\documentclass{emulateapj}
\shorttitle{Nearby Analog of $z\sim$2 Compact Galaxies}
\shortauthors{Jiang et al.}
\begin{document}
\title{A Nearby Analog of $z\sim$2 Compact Quiescent Galaxies with a Rotating Disk}
\author{Fangzhou Jiang\altaffilmark{1}, Pieter van Dokkum\altaffilmark{1}, Rachel Bezanson\altaffilmark{1} and Marijn Franx\altaffilmark{2}}
\altaffiltext{1}{Department of Astronomy, Yale University, New Haven, CT 06511, USA;}
\altaffiltext{2}{Leiden Observatory, Leiden University, P. O. Box 9513, NL-2300 RA Leiden, The Netherlands}
\email{fangzhou.jiang@yale.edu}

\slugcomment{Accepted for publication in ApJ Letters}

\begin{abstract}
Recent studies have identified a population of compact quiescent galaxies at $z\sim2$. These galaxies are very rare today and establishing the existence of a nearby analog could allow us to study its structure in greater detail than is possible at high redshift. Here we present such a local analog, \objectname{NGC 5845}, which has a dynamical mass of $M_{\tiny \textrm{dyn}}=4.3\pm 0.6\times 10^{10}M_{\sun}$ and an effective radius of only $r_{\tiny \textrm{e}}=0.45\pm0.05$kpc.
We study the structure and kinematics with {\it HST}/WFPC2 data and previously published spatially resolved kinematics.
We find that \objectname{NGC 5845} is similar to compact quiescent galaxies at $z\sim2$ in terms of size versus dynamical mass ($r_{\tiny \textrm{e}}$-$M_{\tiny \textrm{dyn}}$), effective velocity dispersion versus size ($\sigma_{\tiny \textrm{e}}$-$r_{\tiny \textrm{e}}$), and effective velocity dispersion versus dynamical mass ($\sigma_{\tiny \textrm{e}}$-$M_{\tiny \textrm{dyn}}$).  
The galaxy has a prominent rotating disk evident in both the photometry and the kinematics: it extends to well beyond $\ga 1/3$ effective radius and contribute to $\ga 1/4$ of the total light of the galaxy. Our results lend support to the idea that a fraction of $z\sim2$ compact galaxies have prominent disks and positive mass-to-light ratio gradients, although we caution that \objectname{NGC 5845} may have had a different formation history than the more massive compact quiescent galaxies at $z\sim2$.
\end{abstract}


\section{Introduction} \label{intro}
In the local universe, massive galaxies are primarily red, dead elliptical galaxies. Studies have uncovered a population of massive, quiescent galaxies as early as $\sim$3Gyr after the big bang (e.g., \citealp{Labbe05, Kriek06, Williams09}).These galaxies are structurally very different from non-star forming galaxies in the nearby universe: quiescent galaxies with stellar mass of $\sim10^{11}M_{\sun}$ at $z\sim2$ are much more compact than galaxies of similar mass at $z=0$, with effective radii a factor of $\sim$3-5 smaller and average stellar densities a factor of $\sim$180 higher than their low-redshift counterparts (e.g., \citealp{Daddi05, Trujillo06, Trujillo07, Toft07, 
vD08, Cimatti08, Franx08, 
Stockton08, Williams10}). Although there can be systematic uncertainties in determining the compactness, such as underestimation of effective radii due to either low signal-to-noise ratio (S/N), and radial mass-to-light ratio ($M_{\star}/L$) gradients, recent studies employing image stacks (e.g., \citealp{vD10}) and very deep individual images (e.g., \citealp{Szomoru10}) have further confirmed their compactness.
Minor mergers may play the dominant role in causing compact $z\sim2$ galaxies to grow "inside-out" (e.g., \citealp{Bezanson09, Hopkins09, vD10}), i.e., the galaxies gradually built up their outskirts over the past $\sim$10Gyr while the mass within a certain core radius is nearly constant with redshift. 

A key issue is the structure and kinematics of the high redshift compact galaxies. The structure reflects the formation mechanism: a major merger may lead to a bulge-like, non-rotating system whereas a dissipative collapse may lead to a rapidly rotating galaxy. Some $z\sim2$ quiescent galaxies have high ellipticity (e.g., van Dokkum et al. 2008, Stockton et al. 2008, van der Wel et al. 2011) but others are quite round (van der Wel et al. 2011), and distinguishing a population of pure disks from a population of bulges is difficult based on photometry alone for these $z\sim2$ quiescent galaxies. 

Such arguments have prompted searches for nearby analogs of compact quiescent galaxies (e.g., Stockton et al. 2010 for $z=0.5$ - $0.8$; Taylor et al. 2010 for local galaxies). It turns out that compact quiescent galaxies in the local universe are very rare. Selection effect alone, namely the incompleteness of Sloan Digital Sky Survey (SDSS) for high surface brightness nearby galaxies, can not account for this dearth (e.g., \citealp{Franx08, Taylor10}). 

In this Letter, we identify a local compact galaxy from Faber et al. (1989) and describe its stellar and dynamical structure, and present it as a potential analog to compact galaxies at $z\sim2$.  

\section{A Compact Galaxy at $z=0$: NGC 5845} \label{selection}

\begin{figure}
\epsscale{1.}
\plotone{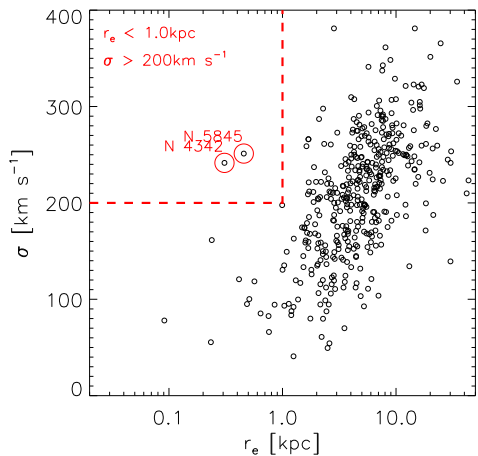}
\caption{Selection of NGC 5845 as a compact galaxy from the Seven Samurai Survey (e.g., \citealp{Faber89}). The whole sample is plotted using black open circles. Given the selection criteria $r_{\tiny \textrm{e, deV}} < 1.0$kpc and velocity dispersion $\sigma > 200$km s$^{-1}$ (the red box), only NGC 5845 and NGC 4342 stand out as outliers.\label{SevenSamurai}}
\end{figure}

\objectname{NGC 5845} and \objectname{NGC 4342} stand out in the $\sigma$-$\log r_{\tiny \textrm{e, deV}}$ diagram (Figure \ref{SevenSamurai}) as the only two objects with $r_{\tiny \textrm{e, deV}}<1$kpc and $\sigma>200$km s$^{-1}$ in the sample of $\sim 400$ nearby quiescent galaxies (Faber et al. 1989). The size cut and the sigma cut is to ensure that the galaxies are dynamically comparable to those of the $z\sim2$ compact galaxies. We present a case study on \objectname{NGC 5845} in the context of comparison to the $z\sim2$ compact quiescent galaxies and discuss the less luminous \objectname{NGC 4342} in Section \ref{discussion}.
\objectname{NGC 5845} is a nearby early-type galaxy in the \objectname{NGC 5846} group, one of the \objectname{Virgo} III groups on the east side of the \objectname{Virgo} cluster (\citealp{Mahdavi05}; \citealp{Eigenthaler10}). 

We characterize the kinematics of \objectname{NGC 5845} using the line-of-sight velocity dispersions (LOSVDs) within a given aperture tabulated in \citet{Gebhardt03}, and the two-dimensional (2D) kinematics maps from the SAURON survey archive. 

\begin{figure*}
\epsscale{0.8}
\plotone{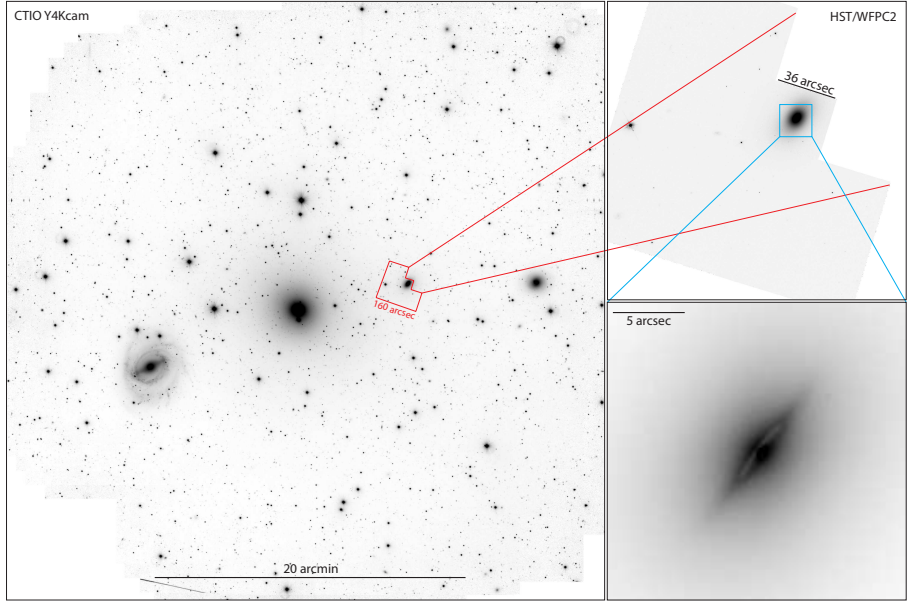}
\caption{Image of NGC 5846 group and NGC 5845. Left: the $V$ band CTIO image of the NGC 5846 Group. Upper right panel: the $V_{\tiny \textrm{F555W}}$ band {\it HST}/WFPC2 image with NGC 5845 in the PC1 aperture. Lower right: the zoomed-in view of NGC 5845 cropped from the $V_{\tiny \textrm{F555W}}$ band {\it HST}/WFPC2 image. A dust disk and a prominent stellar disk can be seen in the center. Gray scales are adjusted respectively. \label{images}}
\end{figure*}

Another key to understanding the structure of \objectname{NGC 5845} is to fully characterize its light profile out to large radii. We include WFPC2 imaging to probe the inner regions of the galaxy and deep ground-based imaging to constrain the outskirts.
$V_{\tiny \textrm{F555W}}$ and $I_{\tiny \textrm{F814W}}$ images are retrieved from the public archive of {\it HST}/WFPC2. Multiple exposures are combined using the CRREJ task within IRAF. The total exposure time is 2140s in $V_{\tiny \textrm{F555W}}$ and 1120s in $I_{\tiny \textrm{F814W}}$. The sky is subtracted from the PC1 chip by extrapolating from the WF chips. 
A very deep exposure centered on \objectname{NGC 5846} of more than 17 hr in the $V$ band is obtained with the CTIO Yale 1 m Telescope using Y4KCam (left panel of Figure \ref{images}), following the observing strategy of the OBEY survey (Tal et al. 2009). 

The WFPC2 image shows that the galaxy has a prominent stellar disk, as well as a smaller dust disk. They have been described as "nuclear" disks (e.g., van Dokkum \& Franx 1995) but as we will show later: the stellar disk is prominent and extends well beyond $\ga1/3$ effective radius of the galaxy. The morphology of the galaxy is usually given as "E" (e.g., Kormendy et al. 1994) but this is influenced by the resolution of ground-based observations.

\section{Comparison to Compact Massive Quiescent $z\sim2$ Galaxies} \label{comparison}

We first ask the question whether \objectname{NGC 5845} is similarly compact as quiescent galaxies at $z\sim2$, so we measure the size, velocity dispersion, and mass of it and compare them to those of high-redshift compact quiescent galaxies.

\begin{figure*}
\epsscale{1.}
\plotone{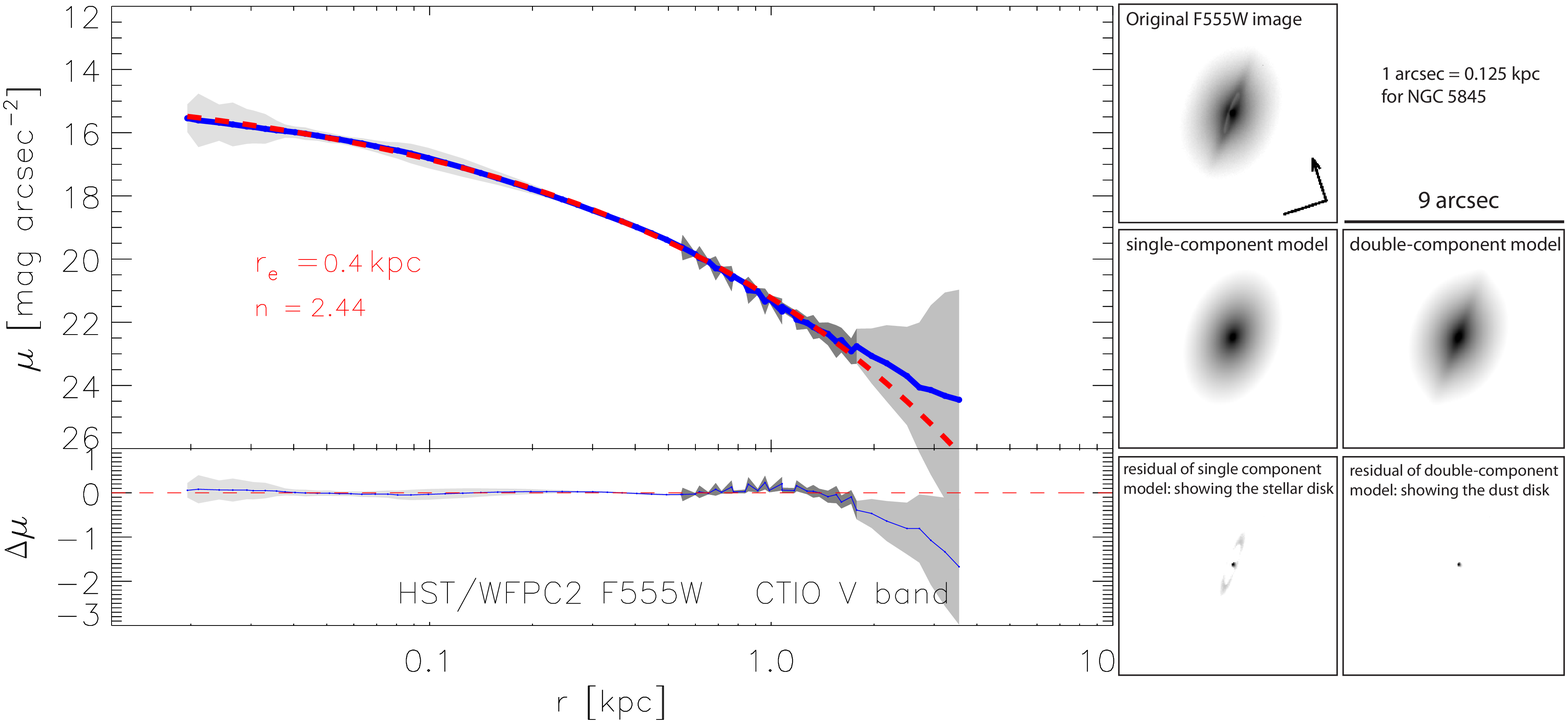}
\caption{ Size of NGC 5845. Left: 1D surface brightness profile (thick blue curve) and the residual (thin blue curve) of the 1D S\'{e}rsic fit (thick dashed red curve) -- the shaded areas indicating the ranges of the two data sources (WFPC2 and CTIO: light gray; overlapping: dark gray) and exaggerated errors (40 times and 15 times respectively for the profile and the residual). The bright tail of the profile is due to the central galaxy NGC 5846. Right: the models and residuals of 2D fits -- the double-S\'{e}rsic fitting constrains the stellar disk component better than a single-component model. The gray scale ranges are set to the same in all the images to enable comparison between one-component and two-component fits. The bright centers in the residual images indicate the dust disk; the dark area around the dust disk is the stellar disk.  \label{size}}
\end{figure*}

First, we determine the half light radius $r_{\tiny \textrm{e}}$ from both 1D and 2D fits -- the 1D algorithm provides more information about the radial variation of the structural parameters of the galaxy, which are essential in helping us to determine the inner structure of the galaxy; while the 2D algorithm is the method used at high redshift.  
We run the ELLIPSE task in IRAF to fit isophotes to both the very deep CTIO $V$ band image and the $V_{\tiny \textrm{F555W}}$ band {\it HST}/WFPC2 image -- SExtractor is used to mask the crowded CTIO image; and for the $V_{\tiny \textrm{F555W}}$ image, we interactively mask out the centermost dust disk according to the $V_{\tiny \textrm{F555W}}/I_{\tiny \textrm{F814W}}$ map. The circularized radial surface brightness profile is shown in Figure \ref{size}. 
In the best-fit 1D S\'{e}rsic model (\citealp{Sersic63}), the effective radius is $r_{\tiny \textrm{e}}=0.40\pm0.04$kpc where the error is dominated by the uncertainty in distance measurement $25.9\pm2.7$Mpc (Tonry et al. 2001), and the S\'{e}rsic index is $n=2.44$. Our 1D size is slightly smaller than 0.51kpc by \citet{Soria06} (in $R$ band), and 0.56kpc by \citet{Trujillo04} (also in $V$ band), both of which are based on the same distance measurement used here. S\'{e}rsic index varies amongst different published measurements from 2.44 (this work) to 2.82 (\citealp{Trujillo04}) and 3.22 (\citealp{Graham01}). 
GALFIT (\citealp{Peng10}) is a standard tool in probing the size and morphology of high-redshift galaxies. We utilize this 2D fitting code for only the higher resolution $V_{\tiny \textrm{F555W}}$ image. A single S\'{e}rsic component gives $r_{\tiny \textrm{e}}=0.49\pm0.05$kpc and $n=2.35$, which agrees well with those derived from the 1D profile. If we fit with double S\'{e}rsic components, the inner disk component with S\'{e}rsic index fixed to $n=1.00$ and the outer bulge component with free S\'{e}rsic index in order to attempt to perform a simple bulge-disk decomposition, then we get $r_{\tiny \textrm{e,bulge}}=0.51\pm0.05$kpc ($n=2.27$) and $r_{\tiny \textrm{e,disk}}=0.16\pm0.02$kpc. 
Taking into account the three size measurements, we will adopt $r_{\tiny \textrm{e}}=0.45\pm0.05$kpc and $ n\approx 2.4$ in the following analysis.

Second, we constrain the mass of \objectname{NGC 5845}.
Similarly to what applies to high-redshift compact quiescent galaxies, the dynamical mass is derived using size, S\'{e}rsic index, and line-of-sight luminosity-weighted velocity dispersion within effective radius:
\begin{equation}\label{dyn1}
M_{\tiny \textrm{dyn}}=\frac{ \beta(n) \sigma_{\tiny \textrm{e}}^2 r_{\tiny \textrm{e}} }{G},
\end{equation} 
where $\beta(n)=8.87-0.831n+0.0241n^2$ (\citealp{Bertin02}). We use the long-slit kinematics along the major axis (Gebhardt et al. 2003) and fold the data in half to derive the radial LOSVD profile $\sqrt{v^2+\sigma^2}(r)$, where $v$ and $\sigma$ are the first and second order Gauss-Hermite velocity moments. Rather than applying aperture corrections we can calculate the luminosity-weighted velocity dispersion within $r_{\tiny \textrm{e}}$ directly from the radial kinematic profile, which gives $\sigma_{\tiny \textrm{e}}=258\pm13$km s$^{-1}$, where the error of 5\% is adopted to allow for systematics. Our result is slightly larger than $\sigma_{\tiny \textrm{e}} =239$km s$^{-1}$ (Cappellari et al. 2006). 
Using $r_{\tiny \textrm{e}}=0.45\pm0.05$kpc and $n=2.44$, we get $M_{\tiny \textrm{dyn}}=4.3\pm0.6\times 10^{10}M_{\sun}$ from Equation (\ref{dyn1}). 

Stellar mass $M_{\star}$ is derived by fitting stellar population synthesis (SPS) models to photometric redshifts, which also constrain the stellar age, dust content, star formation timescale, metallicity, and star formation rate (SFR).
The spectral energy distribution (SED) consists of SDSS $u, g, r, i, z$ and 2MASS $J, H, K$ broadband photometries, all corrected for Galactic extinction according to \citet{Schlegel98}. We run the SPS code FAST (\citealp{Kriek09a}) with the stellar population templates by \citet{BC03}, assuming an exponentially declining star formation history, \citet{Chabrier03} IMF, and \citet{Calzetti00} reddening law. 
The SED is best fit with a stellar mass of $M_{\star}=1.20^{+0.70}_{-0.44}\times10^{10}M_{\sun}$, with $\tau = 1.0$Gyr, age$=7.9^{+0.2}_{-5.4}$Gyr, SFR$=0.01M_{\sun}$yr$^{-1}$, and $A_{V}=0.3$, where an error of $\sim0.2$ dex in $M_{\star}$ is allowed to account for systematic uncertainties (e.g., \citealp{Conroy09}).

\begin{figure*}
\epsscale{1.}
\plotone{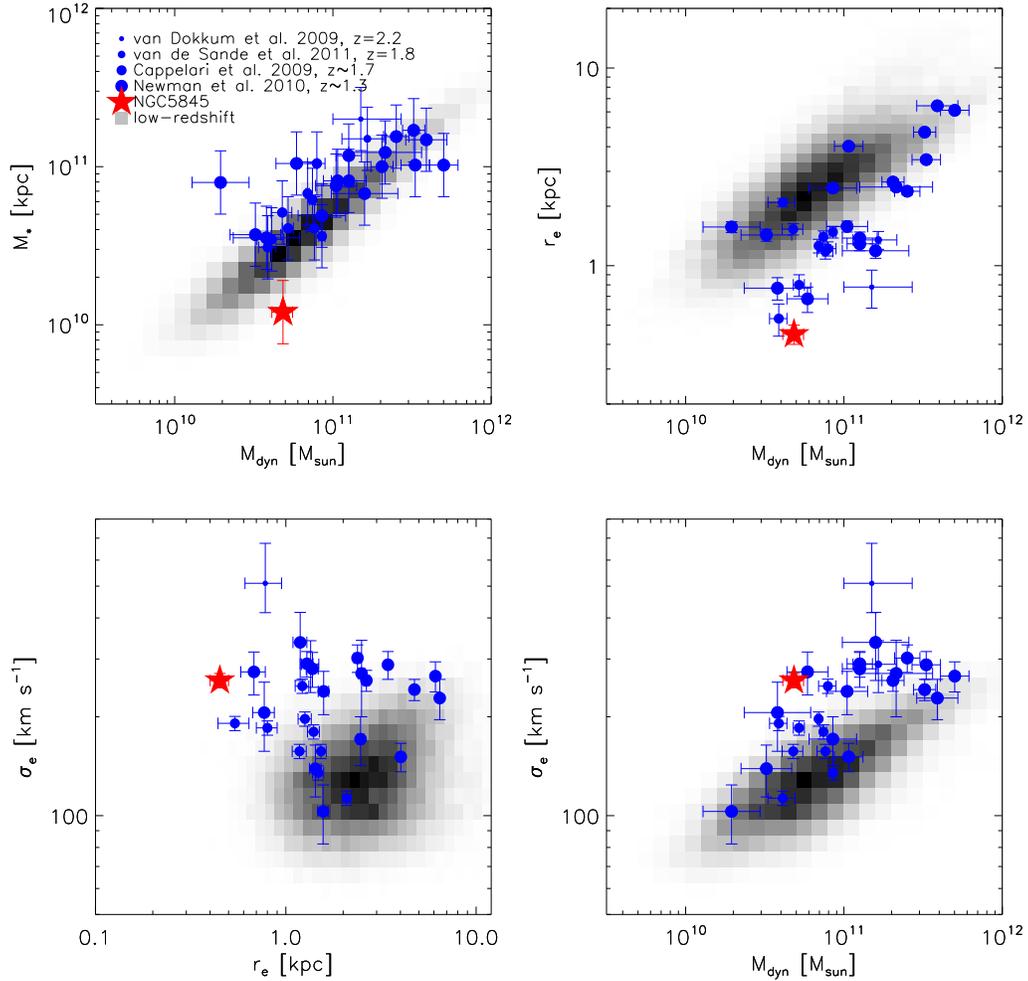}
\caption{Comparison of NGC 5845 (red star) to high-redshift compact galaxies (blue filled circles) and quiescent low-redshift SDSS galaxies (gray cloud). Upper left: stellar mass versus dynamical mass. Upper right: size versus dynamical mass; NGC 5845 seems to fall in the high-redshift size-mass sequence, which has an obvious offset to the nearby sequence. Lower left: velocity dispersion versus size; low-redshift galaxies with a similar velocity dispersion are a factor of $\sim$5 larger in size than NGC 5845. Lower right: velocity dispersion versus dynamical mass; the velocity dispersion is a factor of $\sim 2$ higher than low-redshift galaxies of similar mass. \label{diagrams}}
\end{figure*}

We are now in a position to compare this nearby galaxy to $z\sim2$ quiescent galaxies.
In Figure \ref{diagrams} we compare the sizes, masses and velocity dispersions of \objectname{NGC 5845}, high-redshift compact galaxies, and low-redshift SDSS galaxies ($0.05<z<0.07$; \citealp{York00}). The selection of the low-redshift sample follows \citet{vdS11}. The stellar masses are all converted to the \citet{Chabrier03} IMF, and dynamical masses are all derived using Equation (\ref{dyn1}). 
Generally, \objectname{NGC 5845} has obvious offsets from the low-redshift population. In particular, the very high velocity dispersion makes \objectname{NGC 5845} a prominent outlier of low-redshift galaxies in not only the $\sigma_{\tiny \textrm{e}}$-$r_{\tiny \textrm{e}}$ diagram similar to our selection diagram but also the $\sigma_{\tiny \textrm{e}}$-$M_{\tiny \textrm{dyn}}$ diagram: its velocity dispersion is a factor of $\sim$2 higher than similarly massive low-redshift galaxies; and low-redshift galaxies with similar velocity dispersion are a factor of $\sim$5 larger than \objectname{NGC 5845}. \objectname{NGC 5845} is similar to the high-redshift population in terms of $\sigma_{\tiny \textrm{e}}$-$r_{\tiny \textrm{e}}$, $\sigma_{\tiny \textrm{e}}$-$M_{\tiny \textrm{dyn}}$, and $r_{\tiny \textrm{e}}$-$M_{\tiny \textrm{dyn}}$, but seems to have a lower $M_{\tiny \textrm{dyn}}/M_\star$ ratio. The clearest offsets from the $z=0$ data is seen in the $\sigma_{\tiny \textrm{e}}$-$r_{\tiny \textrm{e}}$ plane. Here \objectname{NGC 5845} is offset from the local relation by more than 0.5 dex in both axes. We conclude that \objectname{NGC 5845} is an analog of the recently identified population of compact, quiescent high redshift galaxies at $z\sim2$. 

\section{A Rotating Disk}\label{disk}

\begin{figure*}
\epsscale{1.}
\plotone{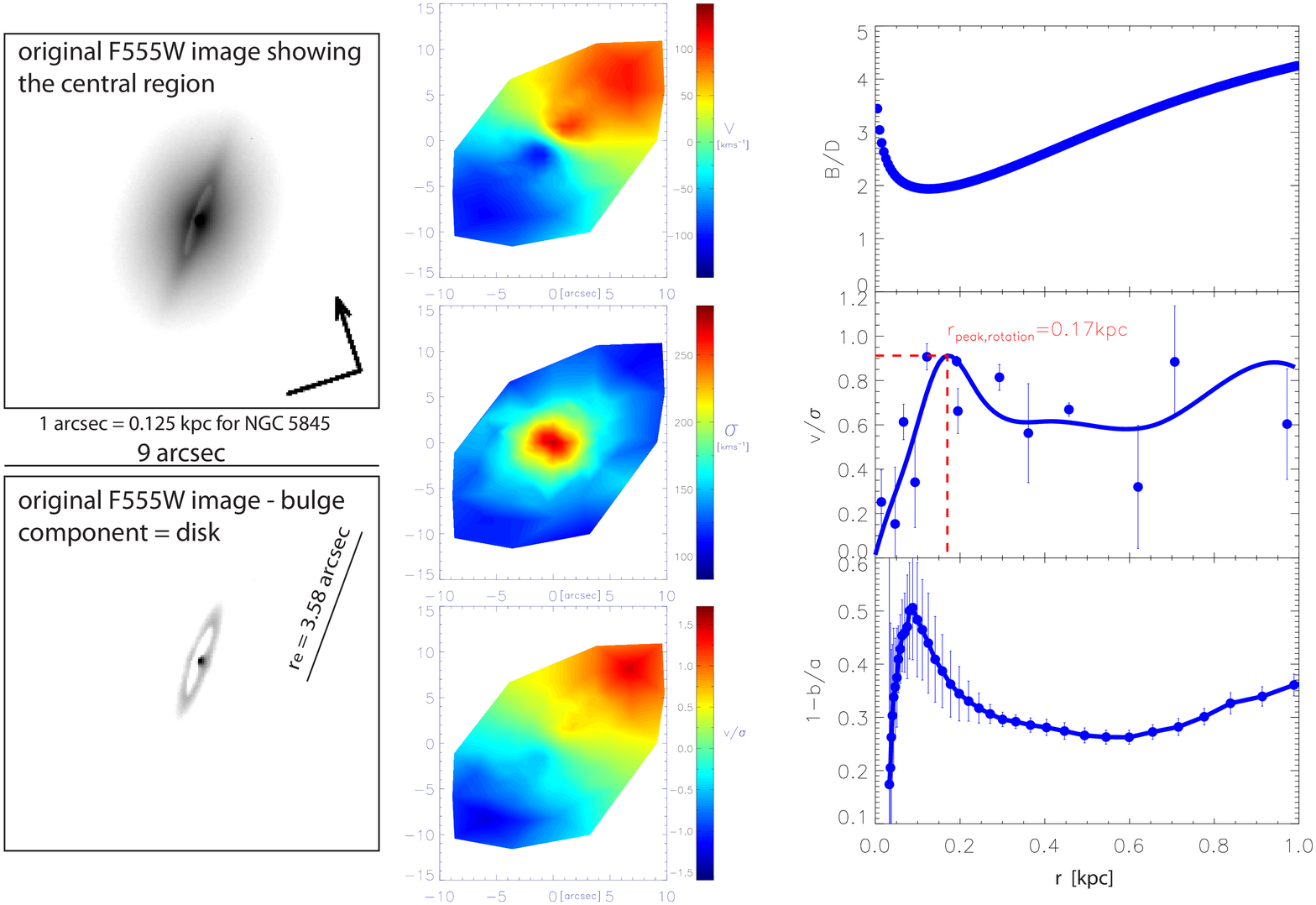}
\caption{Rotating disk in NGC 5845. In the right column, the blue filled dots with error bars are the $v/ \sigma$ profile (with the blue curve as a cubic spline fit) and ellipticity profile according to the kinematics by Gebhardt et al. (2003) and our photometry. The middle column shows the SAURON kinematics maps, from which we can tell that $v$ and $v/\sigma$ both have an inner peak and an outer plateau, consistent with the $v/ \sigma$ profile.  $r_{\tiny \textrm{peak}}$ is defined as the radius at which rotational velocity is comparable to random motion, and the dip in the bulge/disk ratio profile match the rotation peak and the peak in ellipticity, which all constrain the extent of the disk. The disk extends to $\ga r_{\tiny \textrm{e}}/3$ and contributes to $\ga 1/4$ the total luminosity. The images: the residual with the bulge component subtracted from the original image reveals the disk; the extent of the disk estimated from photometry is consistent with the radius of rotation peak. \label{disk&profiles}}
\end{figure*}

A prominent central disk is already revealed in the lower right panel of Figure \ref{images}, and
we wish to estimate the size of the disk and its contribution to the total luminosity of the galaxy quantitatively. The radius at which $v/ \sigma$ approaches an inner peak near unity is defined as the extent of the central disk $r_{\tiny \textrm{peak}}$. We use the kinematics by Gebhardt et al.(2003) to estimate the radius and show the 2D kinematics maps from SAURON survey. As shown in Figure \ref{disk&profiles}, $r_{\tiny \textrm{peak}}\approx 0.17$kpc is consistent with the position of the dip in the B/D profile, indicating that the disk dominate at least the inner $\ga r_{\tiny \textrm{e}}/3$. This is in good agreement with the half-light radius of the disk component found in the double-S\'ersic fit. The range of high flatness is also  within $r_{\tiny \textrm{peak}}$. 
The light concentration index $C_{r_{\tiny \textrm{e}}}(\alpha)$ defined in \citet{Graham01}, is the ratio of flux inside some fraction $\alpha$ of $r_{\tiny \textrm{e}}$ to the total flux inside $r_{\tiny \textrm{e}}$. For \objectname{NGC 5845}, $C_{r_{\tiny \textrm{e}}}(1/3)\approx 0.42$ implies that the region dominated by the disk component contributes $\ga 1/4$ the total luminosity. The B/D profile
also shows that the contribution to the total light from the bulge and disk is $\sim3:1$, and the disk contributes up to 1/3 the light within $r_{\tiny \textrm{e}}$.
The velocity map and $v/\sigma$ map also show inner peaks and outer plateaus as revealed by the $v/\sigma$ profile.

The disk is better revealed when the model bulge component of the galaxy is subtracted from the original $V_{\tiny \textrm{F555W}}$ band image (lower left panel of Figure \ref{disk&profiles}).

\section{Discussion}\label{discussion}

In this work, we have illustrated the compactness of \objectname{NGC 5845} and investigated its prominent rotating disk. The size of the galaxy is unusually small given its dynamical mass, which makes it an outlier with respect to the nearby quiescent galaxies and similar to $z=1.5$-$2.5$ compact quiescent galaxies. We first ask whether our results might have been expected given the observed scatter in the size-mass relation in the SDSS survey.
We fit a size-mass relation to the SDSS galaxies  $\log r_{\tiny \textrm{e} } = -5.2 +  0.53\log(M_\star/M_\sun)$, which is consistent with Shen et al. (2003) and van de Sande et al. (2011), and estimated the standard deviations to this relation in 10 stellar mass bins. We find that \objectname{NGC 5845} is $\sim$3$\sigma$ below the local size-mass relation based on the standard deviation in size in the mass bin that it belongs to. Given that the size distribution in each mass bin can be well described by a log-normal distribution (e.g., Shen et al. 2003), we expect to find 0.54 galaxies as compact as NGC 5845 in 400 local SDSS galaxies. Hence, the probability of finding 1 in the Faber et al. (1989) sample is $\sim 40$\,\%, and the probability of finding 2 is $\sim 10$\,\%. In other words, finding 1 or 2 of these objects in our "backyard" could have been expected. The presence of a rotating disk dominating the inner $\ga1/3$ effective radius of this nearby analog suggests that such structure could also be present in high-redshift counterparts, consistent with the high flattening of some $z\sim2$ compact galaxies (e.g., van Dokkum et al. 2008, van der Wel et al. 2011). 

The question follows naturally whether this local compact quiescent galaxy is a survivor of high redshift counterparts or a newly formed analog. We approach this question in two ways: by looking at its environment and by estimating its stellar age. First, \objectname{NGC 5845} is situated in the 
\objectname{NGC 5846} group, the third most massive aggregate of early-type galaxies (only after \objectname{Virgo} and \objectname{Fornax}) in the local universe (\citealp{Eigenthaler10}), at a distance of $\sim$65kpc from the central giant elliptical galaxy (Figure \ref{images}). Interestingly, two other early-type galaxies in this group, \objectname{NGC 5846A} and \objectname{NGC 5846-205}, are also known to be compact: Mahdavi et al. (2005) almost excluded \objectname{NGC 5845} together with \objectname{NGC 5846A} and \objectname{NGC 5846-205} as group candidates in their study of \objectname{NGC 5846} group because of their relatively high surface brightness and small size. These two galaxies are even closer to the central galaxy: both are blended with the outskirts of \objectname{NGC 5846}. A similar situation exists near \objectname{M87}: \objectname{NGC 4486A} is also compact in terms of size (only $\sim$0.3kpc) and $\sigma$ ($\sim$240km s$^{-1}$, e.g., \citealp{Prugniel11}). Additionally, \objectname{NGC 4486A}  is similar to \objectname{NGC 5845} in some other aspects. They are both fast rotators: \objectname{NGC 4486A} reaches $v_{\tiny \textrm{max}}/\sigma_{0}\approx 1$. They both have a dust disk and stellar disk (\citealp{Kormendy01}). The proximity of these four galaxies to group centers and their similarities suggest that they might be tidally stripped or undergoing tidal stripping. 
However, \objectname{NGC 5845} stands out as a much more massive galaxy compared with the other three, and it is also further away from the group center. We also note that \object{NGC 4342}, the other galaxy in the Faber et al. sample with $r_{\tiny \textrm{e, deV}}<1$kpc and $\sigma>200$, also has a prominent disk (e.g., Cretton \& van den Bosch 1999). However, \objectname{NGC 4342} is relatively faint at $M_B = -17.5$ compared to \objectname{NGC 5845} at $M_B = -18.7$, so when placed at $z\sim 2$ it would not likely be sufficiently bright to be in samples of quiescent high redshift galaxies.
Second,
we infer that if the stellar age of \objectname{NGC 5845} is significantly older than its high-redshift counterparts, it could have been a normal-looking quiescent galaxy (i.e., a descendant of a massive compact quiescent galaxy) without tidal stripping: tidal stripping might help it retain the progenitor's structural appearance. If this is the case, we expect some imprints in metallicity. The metallicity is indeed very high for its $M_\star / L$: [$Z$/H] $= +0.05$ within $r_{\tiny \textrm{e}}$ and [$\alpha$/Fe] $= +0.24$ (Kuntschner et al. 2010), in the range of giant ellipticals. If its stellar age is similar to the high-redshift compact galaxies ($\approx$0.5-2.5Gyr; e.g., \citealp{Kriek09b}), \objectname{NGC 5845} could be at a similar evolutionary stage as the high-redshift counterparts, namely being a newly formed analog. Either option, if grounded with further evidence, would be supplementary to the "inside-out" growth scenario.
However, there are discrepancies in the stellar age estimates, with both younger results for instance, $2.5^{+0.4}_{-0.5}$Gyr (\citealp{Soria06}) and older results $8.9^{+2.6}_{-2.0}$Gyr (\citealp{McDermid06}), $11.2^{+0.5}_{-1.0}$Gyr (\citealp{Kuntschner10}) reported. Our SED fitting gives $\sim$3-8Gyr, but we note that the galaxy does not have a uniform stellar population, as it is almost certainly forming stars in the central dust disk.

Further complicating matters is that the galaxy also has ongoing
    star formation, in its central dust disk (Soria et al. 2006; Shapiro
   et al. 2010). This may suggest that the stellar disk is still
  forming, and is younger than the rest of the galaxy. If this is
   the case, the radial age gradient would imply a positive $M_\star/L$
   gradient, which could result in erroneous size measurements
   (e.g., Hopkins et al. 2009).

We have demonstrated that detailed studies within the effective radii of compact massive galaxies are needed to establish their structure and constrain their formation histories. Unfortunately, were \objectname{NGC 5845} placed at $z\sim 2$, its entire disk would fall within a single WFC3 pixel! Progress may come from other nearby examples, or from observations of lensed high redshift galaxies. A definitive test of the structure of the $z\sim2$ quiescent galaxies requires the measurement of spatially resolved kinematics. This may be possible for very bright examples, although for the general population we may have to await the next generation of telescopes.

\acknowledgments
We wish to thank Tomer Tal at Yale University for helpful discussions.


\end{document}